\documentclass[fleqn,usenatbib]{mnras}

\usepackage{newtxtext,newtxmath}

\usepackage[T1]{fontenc}
\usepackage{lscape}
\DeclareRobustCommand{\VAN}[3]{#2}
\let\VANthebibliography\thebibliography
\def\thebibliography{\DeclareRobustCommand{\VAN}[3]{##3}\VANthebibliography}

\usepackage{graphicx}	
\usepackage{amsmath}	

\title[Broad-band X-ray spectra of faint AGN ]{Dealing with broad-band X-ray spectra of faint AGN: a case study}

\author[Molina et al.]{
M. Molina$^{1}$\thanks{E-mail: manuela.molina@inaf.it}
A. Malizia,$^{2}$
and L. Bassani$^{2}$
\\
$^{1}$IASF/INAF, via Corti 12, I-20133 Milano, Italy\\   
$^{2}$OAS/INAF, via Gobetti 101, I-40129 Bologna, Italy\\   
}

\date{Accepted 2023 October 17. Received 2023 October 11; in original form 2023 July 27
}

\pubyear{2015}

\begin{document}
\label{firstpage}
\pagerange{\pageref{firstpage}--\pageref{lastpage}}
\maketitle

\begin{abstract}
In this work we analyse 3 average-luminosity hard X-ray selected AGN: ESO 506-G27, IGR J19039+3344 and NGC 7465. 
They have simultaneous Swift/XRT and NuSTAR data never published before and have been poorly studied at X-ray energies. 
These sources make for interesting targets both from a methodological and scientific point of view. Scientifically, 
they are of interest since they are possibly heavily absorbed objects, belong to a peculiar class and are 
variable both in flux and in spectral shape. Methodologically, because it is an interesting exercise to understand 
how existing spectral models can be applied to faint sources and how the use of NuSTAR data alone and then simultaneous and/or 
average data impacts on the spectral parameters determination. In this work we demonstrate
that simultaneous data are not sufficient if their statistical quality is poor. Moreover, we show that also the use 
of time-averaged data when dealing with faint AGN does not always provide confident results as for brighter AGN. 
Regardless of the poor data quality employed in our analysis, we are able to provide insights into the spectral 
characteristics of each source. We analyse in detail for the first time the iron line complex of ESO 506-G27, 
finding not only the presence of the iron K$\alpha$ and K$\beta$ lines, but also of the iron K edge around 7 keV 
in the NuSTAR data. We also highlight changes in the absorption properties of IGR J19039+3344 and 
confirm NGC 7465 to be an unabsorbed type 1 LINER.
\end{abstract}




\section{Introduction}

To study the X-ray spectral properties of Active Galactic Nuclei (AGN), 
broad-band studies are essential. The low energy 
part of the spectrum is important to estimate the absorption intrinsic to the source 
(neutral or ionised, simple or complex)
and to detect the presence of a soft excess. On the other hand, high energy data are 
fundamental to measure the underlying power-law continuum, 
in particular the temperature of the corona, by means of the high energy cut-off, as well as 
the reflection hump around 30 keV (see e.g. \citealt{malizia20}).

A wider energy band has been made available with the advent of the NuSTAR mission \citep{harrison13}; however,
even if the energy coverage is quite broad (3-78 keV), it is not sufficient to properly
measure the soft excess and intrinsic absorption on the low energy part of the spectrum, as well
as the high energy cut-off when it is located above a 100 keV.
In the last few years, several broad-band studies on bright AGN have been published,
either by using NuSTAR data alone (e.g. \citealt{akylas21}) or in combination with data
from soft X-ray observatories, such as Swift/XRT and XMM-Newton, as
well as with higher energy data from either Swift/BAT or INTEGRAL/IBIS. We point out that
both Swift/BAT and INTEGRAL/IBIS provide time-averaged spectra over long 
exposure times, to allow a good signal to noise ratio. These spectra can then be combined with single 
snapshot observations, with the risk of incurring in problematic issues related to source spectral and flux variability. 
Despite the introduction of cross-calibration constants to account for flux variability and/or mismatches in 
the instruments calibration, some level of uncertainty still remains as spectral variability cannot be ruled 
out {\it a priori} 
\citep{lubinski10,fedorova16,Ricci_2017, molina19}. 

For bright ($\sim$10$^{-11}$erg cm$^{-2}$ s$^{-1}$) sources belonging to the
INTEGRAL complete sample, for which simultaneous Swift/XRT 
and NuSTAR were available, 
\citet{molina19} confirmed the results obtained on the same objects but 
using non-simultaneous data by \citet{Malizia:2014}.
In the case of average luminosity AGN (i.e. the large majority 
of the population) and in particular for absorbed objects with spectral complexities (soft
excess, warm absorber, iron line complex), this is not always the case, as it is 
more difficult to characterise their spectral properties; this could be due to a number of reasons, including the non-simultaneity of the data and/or their low statistical quality. 

The INTEGRAL/IBIS and Swift/BAT AGN catalogues (e.g. \citealt{malizia23,Oh18}) list a considerable number of
faint and less well-studied sources, which present some challenges in their analysis, given that their spectra
are quite often characterised by data of poor statistical quality. 

In this work we analyse a set of 3 average-luminosity AGN: ESO 506-G27, IGR J19039+3344 and NGC 7465. They 
have been extracted from the
INTEGRAL AGN catalogues \citep{Malizia_2012, malizia20} and are also detected by BAT \citep{Oh18};
all of them have simultaneous Swift/XRT and NuSTAR data never published before.
All three sources have been poorly studied at X-ray energies, particularly they lack in-depth
analyses of broad-band spectra, except for a study
by \citet{Ricci_2017},  who used average Swift/BAT spectra in combination with non simultaneous soft X-ray 
data.
These sources make for interesting targets both from  a methodological and from a purely scientific point of 
view.
Scientifically, they are of interest since they are possibly heavily absorbed objects, belong to 
a peculiar class of AGN (NGC 7465) and are variable both in flux and in spectral shape on timescales
of months to years. Methodologically, 
because it is an interesting exercise to understand how existing spectral models can be applied
to faint sources and how the use of NuSTAR data alone and then simultaneous and/or average data impacts on the
spectral parameters determination.

The methodology we apply in our study is the following: we first fit single NuSTAR observations on their own; 
then we consider simultaneous XRT/NuSTAR
observations; finally, we analyse average XRT plus 
INTEGRAL/IBIS spectra.
A comparison between the results of the different fitting procedures and the limited works available in the 
literature provides an insight into the various procedures adopted as well as their limitations/improvements. 
As a byproduct of our analysis, we are also able to provide interesting scientific insights on each source, 
despite the constraints imposed by  data quality.

\section{Source selection and Data Reduction}
 The source selected for this analysis are the type 2 active galaxies ESO 506-G27 and IGR J19039+3994
plus  NGC 7465, which has an ambiguous classification between a Seyfert 2 \citep{koss22} and
a low-ionization nuclear emission-line region AGN (LINER, e.g. \citealt{ferruit2000}).
\citet{guainazzi05} measured a high column density (N$_{\rm H}$$\sim$5$\times$10$^{23}$cm$^{-2}$)
thus justifying the type 2 classification, but \citet{ramos09} classified it 
as a type 1 LINER, based on the presence of broad lines in near infra-red spectra.

They have been extracted from 
INTEGRAL AGN catalogues \citep{Malizia_2012, malizia20} 
and have therefore
been firmly detected above 20 keV.
In Table \ref{obs_log} we report the main information for each object (coordinates, source redshift, optical 
classification, Galactic column density, observation details); in the last three columns  of Table \ref{obs_log} we also
include the cleaned exposures and significance of detection as well as the 2-10 keV NuSTAR observed fluxes relative to the 
simultaneous observations (see section \ref{sim_fit}). It is worth noting that all three sources are detected by NuSTAR up to $\sim$70 keV
(see figures relative to the spectral fits).
The three sources have been selected as the target of the present analysis since they 
are potentially peculiar 
sources: they are quite faint in the X-rays and show interesting absorption properties. Besides, they are all
variable, either in flux or in spectral shape and their soft X-ray data are of low statistical quality, as 
well as the NuSTAR ones, which are also below average quality, and have low exposure times.
Therefore the three AGN selected for our scientific/methodological study make for ideal key targets.
Here we report never published before NuSTAR observations, together with simultaneous soft X-ray data
obtained from the XRT telescope on board the Neil Geherels Swift Observatory;
we also analysed a 2006 XMM observation of ESO 506-G27 for the purpose of better characterise the soft part of the spectrum. Time-averaged
spectra from Swift/XRT combined with INTEGRAL/IBIS data are analysed and compare with simultaneous spectra.

XRT data reduction was performed using the standard data pipeline
package (\texttt{XRTPIPELINE} v. 0.13.2) so to produce screened
event files (see e.g. \citealt{landi10}). Source events were extracted 
within a circular region with a radius of 20 pixels (1 pixel corresponding to 2.36 arcsec) 
centred on the source position, while background events were extracted from a source-free
region close to the X-ray source of interest. The spectra were
obtained from the corresponding event files using the \texttt{XSELECT}
v.2.4c software, then binned appropriately using \texttt{grppha}.
We used version v.014 of the response matrices and created 
individual ancillary response files using the task \texttt{xrtmkarf} v.0.6.3.

NuSTAR data (from both focal plane detectors, FPMA and FPMB)
were reduced using the \texttt{nustardas\_04May21\_v2.1.1} and \texttt{CALDB} version
20220118. For our sources, spectral extraction and the subsequent production of
response and ancillary files was performed using the
\texttt{nuproducts} task with an extraction radius chosen depending on the source brightness; 
to maximise the signal-to-noise ratio, background spectra were extracted from 
circular regions of typically 50$^{\prime\prime}$--70$^{\prime\prime}$ radius 
in source-free areas of the detectors. When fitting NuSTAR data a multiplicative
cross-calibration constant has always been taken into account and found to be close
to one in each case, as expected.
All our spectra are background subtracted.

For obtaining the high energy data, 
we used the INTEGRAL IBIS Off-line Scientific Analysis pipeline
(OSA 11.2), which produces spectra, response and ancillary matrices. 

Archival data for all three sources where obtained either through the HEASARC database\footnote{https://heasarc.gsfc.nasa.gov/docs/archive.html} or through the XMM
Science Archive\footnote{https://www.cosmos.esa.int/web/xmm-newton/xsa}.

\begin{table*}
\begin{center}
\caption{Source info and observation log}
\renewcommand{\arraystretch}{1.2}
\resizebox{\linewidth}{!}{%
\begin{tabular}{lccccccccccc}
\hline
{\bf Name}  &   {\bf RA}  & {\bf Dec}& {\bf z} & {\bf Class }&{\bf N$_{\bf \rm H}^{\bf \rm Gal}$} & {\bf Telescope} &{\bf Obs. ID}& {\bf Date}& {\bf Exposure}& {\bf significance$^{\dagger}$}&{\bf F$_{\bf 2-10}$}\\
            &                            &          &         &             & {\bf 10$^{\bf 22}$cm$^{\bf -2}$}&                    & 
        &           &  {\bf ksec}& $\sigma$&{\bf erg cm$^{\bf -2}$s$^{\bf -2}$}\\    
\hline
ESO 506-G27& 12 38 54.59 & -27 18 28.01 & 0.02502 & Sey 2 &0.0534& XRT &00035173002&	15 June 2005& 7.5 & &\\
                &                       &              &         &       &       & XRT & 00035173003 & 15 Aug. 2005 &2.26& &\\
                &                        &              &         &       &       & XRT & 00035173004 & 15 Aug. 2005 &11.5& &\\                
                &                        &              &         &       &       & XRT &00088765001&12 Oct. 2018&1.8&&\\
                &                        &              &         &       &       & XRT &  {\bf 00088765002}&{\bf 26 June 2019} &2.16& 4.23&\\                
                &                        &              &         &       &       & NuSTAR &{\bf 60469006002}& {\bf 26 June  2019}&18.7 &42.34&2.43$\times$10$^{-12}$\\
                &                         &              &         &      &       & XMM&0312191801& 24 Jan. 2006&  7.3& &  \\
                &                        &              &         &       &           & IBIS &  --         &   --         & 227&9.3&\\
IGR J19039+3344 &  19 03 49.14 & +33 50 41.11 & 0.0150 & Sey 2 &0.074& XRT &00090184001 & 26 May 2009 & 10.3& &\\
               &                        &              &         &     &         &XRT& {\bf 00081412001}&{\bf 27 Aug. 2019} &6.0&5.10&\\
                                        &              &         &     &  &       &XRT&00081412002& 3 July 2022 & 4.2 & &\\
                                        &              &         &     &  &       &XRT& 00081412003& 3 July 2022 & 1.4 &&\\
                &                       &              &         &     &       & NuSTAR &{\bf 60161704002}&{\bf 27 Aug. 2019}& 23.5&23.19&9.67$\times$10$^{-13}$\\
NGC 7465       &     23 02 00.96 & +15 57 53.24 & 0.00736 & LINER & 0.0517&XRT & 00011341001 & 9 May 2019& 1.5&& \\
                &                        &              &         &       &       &XRT & {\bf 00081295001}&{\bf 9 Jan. 2020}& 1.6&33.37&\\
                &                        &              &         &       &       &XRT &  00081295002 & 10 Jan. 2020 & 4.8&&\\
                &                        &              &         &       &       & NuSTAR &{\bf 60160815002}&{\bf 09 Jan. 2020}& 20.9&75.44&1.32$\times$10$^{-11}$\\
                &                        &              &         &       &           & IBIS &  --         &  --  & 1253&4.0&\\                
\hline
\multicolumn{12}{l}{$\dagger$: we report the detection significance only for simultaneous observations, while for the total sensitivity achieved when summing the X-ray data used in the time-averaged fits see text.}\\
\multicolumn{12}{l}{Note: IBIS spectra are time-averaged over the whole duration of the mission.}\\
\multicolumn{12}{l}{Note: in bold are highlighted the strictly simultaneous observations.}\\
\multicolumn{12}{l}{Note: 2-10 keV fluxes refer to NuSTAR FPMA data.}\\
\end{tabular}
\label{obs_log}
}
\end{center}
\end{table*}

\section{Broad-band spectral analysis}



\begin{table*}
\begin{center}
\caption{{\bf ESO 506-G27}}
\renewcommand{\arraystretch}{1.2}
\resizebox{\linewidth}{!}{%
\begin{tabular}{lccccccccc}
\hline
    & {\bf N$_{\rm \bf H}$}&{$\bf \Gamma$}&{\bf E$_{\rm \bf Fe}$}&{\bf EW}& {\bf E$_{\rm \bf edge}$}&{\bf Fe Abund.}&{\bf E$_{\rm \bf cut}$}& {\bf R}&{\bf $\chi^{\bf 2}$ (d.o.f.)}\\ 
        &   {\bf 10$^{\bf 22}$ cm$^{\bf -2}$}&    & {\bf keV}&            {\bf eV}&  {\bf keV}& &{\bf keV}&             &         \\
\hline
\multicolumn{10}{c}{\texttt{phabs*[po+zvfeabs*(po+zga)]}}\\
\hline
NuSTAR & 57.78$^{+12.85}_{-11.21}$&1.73$^{+0.14}_{-0.13}$& 6.35$\pm$0.05 & 482$^{+199}_{-153}$&7.11$\pm$0.10 &1.70$^{+0.52}_{-0.44}$ &-- & --&340.09 (355)\\
\hline
\multicolumn{10}{c}{\texttt{phabs*[po+zvfeabs*(pexrav+zga)]}}\\
\hline
NuSTAR  & 62.36$^{+10.86}_{-16.27}$ & 2.04$^{+0.18}_{-0.98}$ &6.35$\pm$0.05& 467$^{+200}_{-153}$&7.10$^{+0.07}_{-0.08}$& 1.35$^{+0.42}_{-0.39}$ & $>$37 & NC& 336.75 (353) \\
\hline
XRT+NuSTAR (sim)& 49.52$^{+7.32}_{-6.56}$&2.04 fixed &6.35$^{+0.05}_{-0.06}$&419$^{+234}_{-142}$&$<$7.17&1.51$^{+0.53}_{-0.37}$&$>$145&$>$0.86 &351.98 (359)\\

\hline
\end{tabular}
}
\label{eso506_fits}
\end{center}
\end{table*}

\begin{table*}
\begin{center}
\caption{IGR J19039+3344}
\renewcommand{\arraystretch}{1.2}
\resizebox{\linewidth}{!}{%
\begin{tabular}{lcccccccc}
\hline
    & {\bf N$_{\rm \bf H}$}& {$\bf \Gamma_{\rm \bf soft}$}&{ $\bf \Gamma$}&{\bf E$_{\rm \bf  Fe}$}&{\bf EW}& {\bf E$_{\rm \bf cut}$}& {\bf R}&{\bf $\chi^{\bf 2}$ (d.o.f.)}\\ 
    &{\bf 10$^{\bf 22}$ cm$^{\bf -2}$}&                               &              &{\bf keV}&{\bf eV}&{\bf keV}&           &                          \\
\hline
\multicolumn{9}{c}{\texttt{phabs*[po+phabs*(po+zga)]}}\\
\hline
NuSTAR & 85.26$^{+78.35}_{-37.67}$ &  1.45$^{+3.89}_{-0.69}$&1.67$^{+1.09}_{-0.32}$& 6.02$\pm$0.11 & 491$^{+2883}_{-342}$& -- & -- & 101.30 (126)\\
\hline
\multicolumn{9}{c}{\texttt{phabs*[po+phabs*(pexrav+zga)]}}\\
\hline
NuSTAR & 73.99$^{+52.01}_{-38.19}$ &$>$0.65& 1.67$^{+0.29}_{-0.21}$ &6.02$^{+0.12}_{-0.11}$& 471$^{+1414}_{-375}$&$>$14 & NC& 99.09 (125) \\
XRT+NuSTAR (sim)& 53.18$^{+28.18}_{-18.20}$&0.83$^{+1.57}_{-0.34}$&1.61$^{+0.37}_{-0.29}$&6.02$^{+0.14}_{-0.12}$&390$^{+588}_{-259}$&68$^{+750}_{-47}$&$>$0.67&110.72(132)\\
\hline
\end{tabular}
}
\label{19039_fits}
\end{center}
\end{table*}

\begin{table*}
\begin{center}
\caption{NGC 7465}
\renewcommand{\arraystretch}{1.2}
\resizebox{\linewidth}{!}{%
\begin{tabular}{lcccccccc}
\hline
    & {\bf N$_{\rm \bf H}$}&{ $\bf \Gamma$}& {\bf E$_{\rm \bf  Fe}$}&{\bf EW}&{\bf E$_{\rm \bf cut}$}& {\bf R}&{\bf c}&{\bf $\chi^{\bf 2}$ (d.o.f.)}\\ 
    &{\bf 10$^{\bf 22}$ cm$^{\bf -2}$}& &{\bf keV}&{\bf eV}& {\bf keV}&  & &\\
\hline
\multicolumn{9}{c}{\texttt{phabs*phabs*(po+zga)}}\\
\hline
NuSTAR   & $<$0.97&1.67$^{+0.05}_{-0.03}$&  6.47$^{+0.08}_{-0.09}$& 146$^{+45}_{-48}$& -- & -- & --&416.24 (446)\\
\hline
\multicolumn{9}{c}{\texttt{phabs*phabs*(pexrav+zga)}}\\
\hline
NuSTAR & $<$2.12 & 1.81$^{+0.19}_{-0.16}$ &6.47$\pm$0.09&126$^{+51}_{-48}$&$>$60 & 1.05$^{+0.90}_{-0.60}$ & --&404.48 (444) \\
\hline
XRT+NuSTAR (sim)&0.58$^{+0.18}_{-0.16}$& 1.80$^{+0.11}_{-0.10}$ &6.48$^{+0.09}_{-0.10}$&127$^{+46}_{-49}$&119$^{+384}_{-54}$ &1.05$^{+0.81}_{-0.57}$ &1.14$^{+0.15}_{-0.12}$ &423.17 (462)\\
AVERAGE (XRT+IBIS) & 0.43$^{+0.12}_{-0.11}$&1.50$^{+0.28}_{-0.12}$& 6.4 fixed & $<$452 &42$^{+142}_{-13}$&$<$2& 0.90$^{+0.68}_{-0.50}$&61.30 (60)\\
\hline
\end{tabular}
}
\label{ngc7465_fits}
\end{center}
\end{table*}

The first step of our study consists in analysing the source spectra 
by fitting NuSTAR data (from both focal plane detectors) on their own. 
Several works in the literature (see for instance \citealt{akylas21}, \citealt{kang20} and
\citealt{kang22}) use this approach, preferring NuSTAR data over
broad-band data obtained from different telescopes and fitted together. 
NuSTAR is able to cover a quite wide energy range which, especially when dealing with faint sources,
can be helpful for the determination of the main continuum shape and to
highlight some spectral features.

Subsequent steps in our analysis
consist in adding soft energy data from simultaneous XRT observations to the NuSTAR ones, 
widening the energy coverage with the intent of better determining the spectral
features, above all those arising at softer energies, such as the soft excess and 
intrinsic absorption.
Lastly, we consider time-averaged 
spectra obtained by combining all available Swift/XRT observations together with high energy 
data from INTEGRAL/IBIS 
with the aim of putting tighter constraints on the high energy cut-off and to compare the
results of these fits with those obtained with simultaneous data.

The source 
spectra were generally binned using \texttt{grppha} in an appropriate way so that
$\chi^{2}$ statistic could be applied. This is always the case for NGC 7465 data (both XRT and NuSTAR),
while for the other two sources this is true for NuSTAR spectra but not 
for XRT single observations, which have very
poor statistics, and data were binned with a lower count value. We have verified that the results
obtained using XRT spectra binned with a minimum of 20 counts per bin do not change
the values of the spectral parameters in the broad-band analysis, therefore, although the Cash statistics
should be adopted in these cases, we nonetheless use the $\chi^{2}$ statistic in the combined XRT/NuSTAR fits
of ESO 506-G27 and IGR J19039+3344.

Spectral analysis was conducted with \texttt{XSPEC} v.12.12.0 \citep{arnaud96}, employing the 
$\chi^2$ statistics; uncertainties are listed at the 90\% confidence level
($\Delta\chi^2$=2.71 for one parameter of interest) and abundances were all set to Solar
with the exception of iron for ESO 506-G27 (see text for details).

\subsection{NuSTAR data}
\begin{figure}
\centering
\includegraphics[scale=0.3]{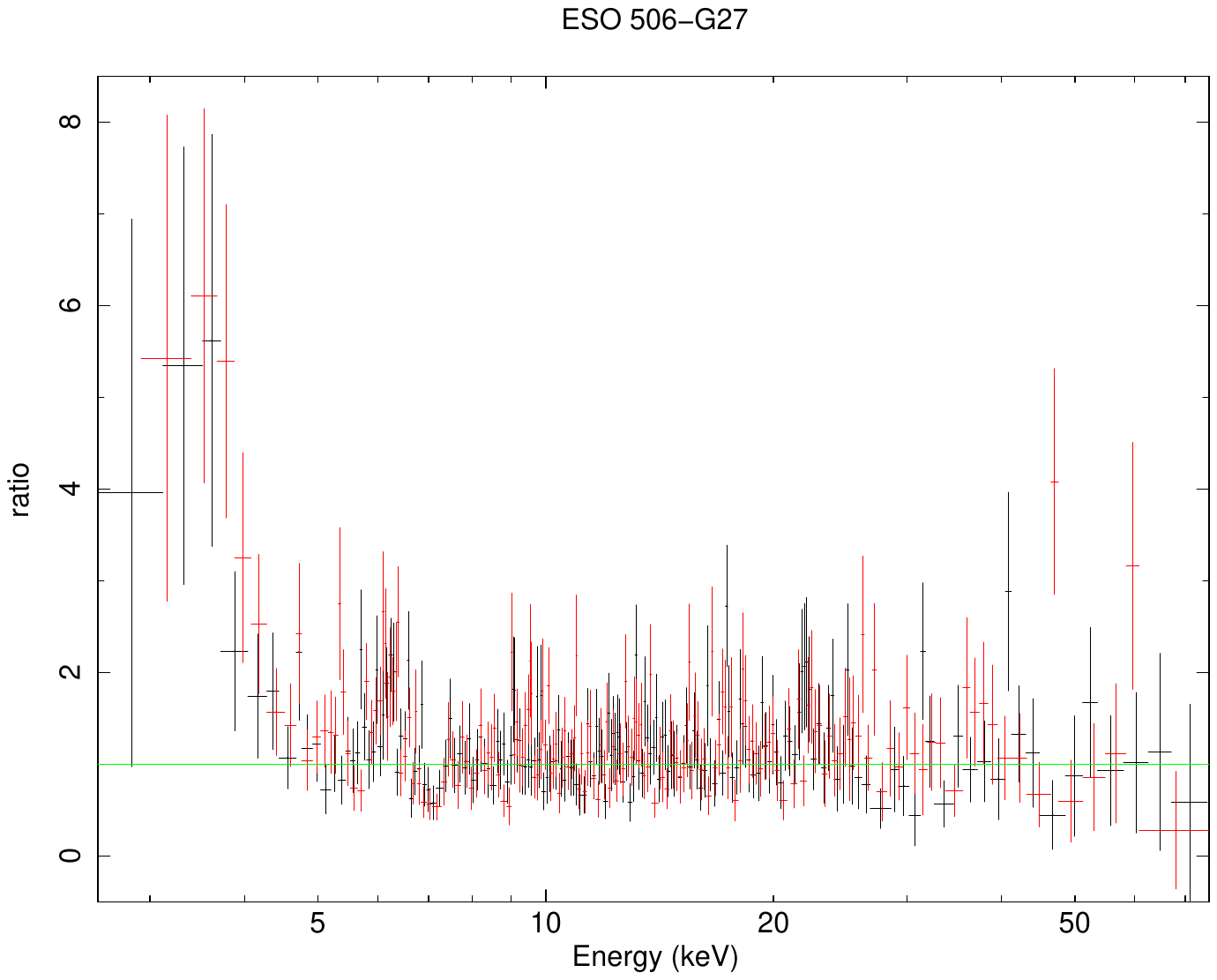}
\caption{{ Model-to-data ratio for ESO 506-G27 (NuSTAR FPMA in black and NuSTAR FMPB in
red). A soft excess component is visible below 5 keV, as well as the iron line complex at around 6.4 keV. }}
\label{eso_ratio}
\end{figure}

\begin{figure}
\centering
\includegraphics[scale=0.3]{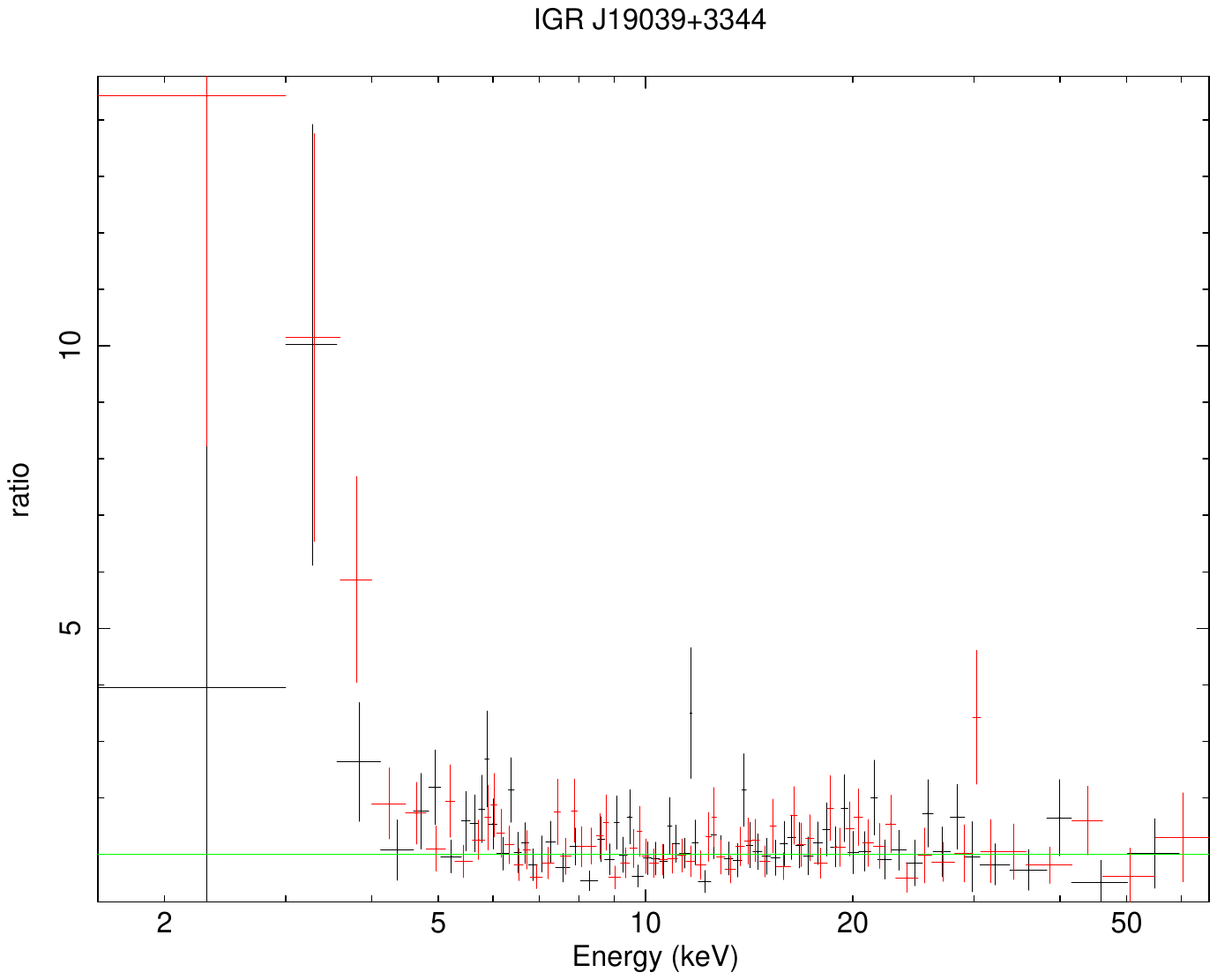}
\caption{{Model-to-data ratio for IGR J19039+3344 (NuSTAR FPMA in black and NuSTAR FMPB in
red). A soft excess component is visible below 5 keV; the iron line is seen at around 6 keV (see text for details).}}
\label{igr19039_ratio}
\end{figure}

We initially fitted NuSTAR data employing a simple power law, absorbed by Galactic \citep{bekhti16}
and intrinsic column densities. A Gaussian component to model the iron line, 
with its width fixed at 10 eV, was also added since residuals are evident around 6 keV in all sources.
The iron line is required in the spectra of all three sources at more than 99\%
confidence level, as determined by employing the $\Delta\chi^2$ test on our fit results.
From these simple fits, it is evident that both ESO 506-G27 and
IGR J19039+3344 show residuals in the low energy part of their spectra 
(see Fig. \ref{eso_ratio} and  Fig. \ref{igr19039_ratio}),
indicative of the presence of a soft excess, 
extending up to almost 5 keV. The presence of this component is further confirmed by
the flattening of the continuum shape
($\sim$1.5 for ESO 506-G27 and $\sim$1.2 for IGR J19039+3344) when
the component is not taken into account. 

We therefore fitted the spectra of ESO 506-G27 and IGR J19039+3344 adding a second power-law to 
our fits to approximately model this component, since NuSTAR data do not cover the energy 
range below 3 keV where most of the soft excess emission is found. The addition of a second, 
soft power-law component to model the soft excess yields an improvement of more than 99.99\% 
for both sources, with ESO 506-G27 having a $\Delta\chi^2$=30.47 for 2 degrees of freedom
(d.o.f.) and IGR J19039+3344 having $\Delta\chi^2$=20.74 (for 2 d.o.f.).
The ESO 506-G27 data are well fitted by tying the two power-law photon
indices (the one describing the soft excess and the one relative to the primary continuum), 
as was commonly done to take into account the scattering component in highly
obscured AGN when dealing with low statistical quality data of highly
(i.e. \citealt{malaguti97}).
In the case of IGR J19039+3344, the best fit is obtained by leaving the two photon
indices independent of one other; this gives a value of the soft power-law 
photon index which is unexpectedly harder than the continuum one. This
could be due to the low statistics at softer energies of our data and as we
already highlighted before, this is but an approximation of the
soft excess component.

We point out that in the case of ESO 506-G27, inspection of the residuals 
suggests that more components are needed to model the iron line, in particular an absorption 
feature at around 7 keV seems to be present (see Fig. \ref{eso_ratio}) likely an
iron K edge given its energy.
We therefore modified the model accordingly (see Table \ref{eso506_fits}), by adding
the \texttt{zvfeabs} component in \texttt{XSPEC}, which describes a redshifted 
photoelectric absorption, where all abundances are tied and set equal to Solar, except for iron.
The model therefore gives as an output the absorbing column density N$_{\rm H}$ as well as the Fe K edge energy.
The introduction of this component yields an improvement of more than
99.99\%, with a $\Delta\chi^2$=21.17 (for 2 d.o.f.).

The baseline models used are 
\texttt{phabs*[po+zvfeabs*(po+zga)]} for ESO 506-G27, \texttt{phabs*[po+phabs*(po+zga)]} for
IGR J19039+3344 and \texttt{phabs*phabs*(po+zga)} for NGC 7465.

Results of these preliminary fits are reported in
the first rows of Tables \ref{eso506_fits}, \ref{19039_fits} and \ref{ngc7465_fits}.  As 
can be seen from the fits for ESO 506-G27 and IGR J19039+3344, after modifying the model to 
account for the soft excess component, we obtained photon indices which are closer
to the canonical one.

Next, NuSTAR spectra have been fitted by substituting in the baseline models the simple absorbed power-law 
with an exponentially cut-off power-law reflected from neutral material
(the \texttt{pexrav} model in \texttt{XSPEC}, \citealt{Magdziarz_1995}) in order to investigate
the high energy cut-off and the reflection component.

The use of the \texttt{pexrav} model yields an improvement 
in the fit of more than 99\% only in
the case of NGC 7465 ($\Delta\chi^2$=11.76, for 2 d.o.f.), whereas for ESO 506-G27  and  IGR J19039+3344 the introduction of the \texttt{pexrav} 
model does not yield  any
improvement. We point out that, despite the \texttt{pexrav} being a phenomenological model,
when the statistics are not particularly good (as is our case), this is the best model to fit the high energy data; 
the use of complex, more physical models can lead to an over-fitting of the data, resulting in 
an incorrect estimate of the spectral parameters, as we will demonstrate in section \ref{borus}.

In all three sources, the high energy cut-off is not constrained and we could
at most find only lower limits on the parameter;  it is worth noting that these lower limits
are quite low with respect to what generally found in other studies (e.g. \citealt{Malizia_2016},
\citealt{molina19}, \citealt{akylas21}).
The reflection fraction is constrained only in NGC 7465 (see Table \ref{ngc7465_fits}), while for the other 
two sources we could not place any constraint on the parameter.

\subsection{Simultaneous broad-band spectral analysis}\label{sim_fit}
Next we took advantage of the available
simultaneous Swift/XRT data covering the soft energy band and fitted the spectra over the
wider 0.5-78 keV energy range. For each source, we employed the baseline models with the \texttt{pexrav} 
instead of the simple power-law and in all fits we added cross-calibration constants to account for mismatches in instrument calibration.
However, in the case of ESO 506-G27 and IGR J19039+3344, the cross-calibration constants between XRT and NuSTAR 
had to be fixed to unity; this choice is mainly motivated by the fact that the Swift/XRT data of ESO 506-G27 and IGR J19039+3344 
are of poor statistical quality (see the detection significance reported in Table \ref{obs_log}) 
and their addition to the NuSTAR data does not allow a better 
determination of the main continuum parameters,
leading to cross-calibration constants not consistent with 1. Indeed, when adding low quality data
(XRT) to medium-to-good quality data (NuSTAR) the goodness of the fit can be diminished,
and in such cases the only way to obtain acceptable fits is to fix the constants to one.
In this way, for the simultaneous data of both sources, we obtain spectral parameters consistent with
those measured with NuSTAR alone. Besides, this is a reasonable choice since the
spectra are strictly simultaneous and therefore the cross-calibration constants are expected to be
consistent with 1. This, however, does not hold for NGC 7465, which has better
quality soft X-ray data than the other two sources; in this case, the cross-calibration constant 
(for simplicity we report only the one between XRT and the NuSTAR FPMA detector) is left 
as a free parameter and its value is consistent (within errors) with unity, as expected
in the case of data taken simultaneously (see third row of Table \ref{ngc7465_fits}).
We point out that in the ESO 506-G27 fit, we had to fix the main continuum power law
photon index to the one obtained by analysing NuSTAR data on their own in order for the
fit to reach convergence. This again is an issue due to the low quality XRT data as well as
to the source spectral complexity.

In general, for all sources the column densities
obtained from the combined XRT/NuSTAR spectra are compatible with those obtained by fitting 
NuSTAR data on their own, suggesting that the addition of poor quality data to average-quality
ones does not impact the fit results. Indeed, in the case of NGC 7465 the absorption
is better constrained than in the single NuSTAR fits, and this is likely due to 
the fact that the XRT data of NGC 7465 are of 
better quality than those of the other two sources (see Table \ref{obs_log}).
 
\begin{figure}
\centering
\includegraphics[scale=0.3]{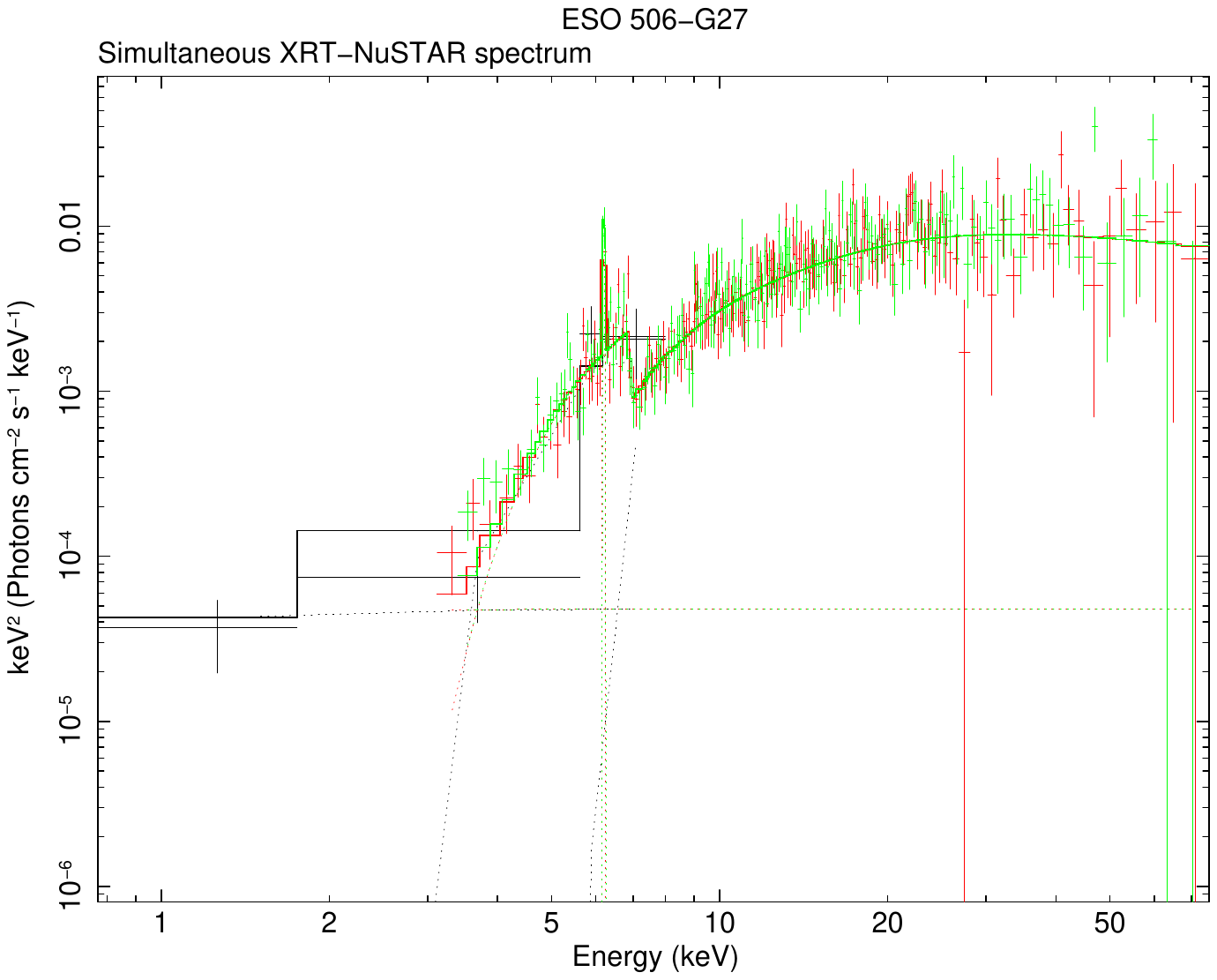}
\caption{{Simultaneous XRT/NuSTAR unfolded spectrum of ESO 506-G27.}}
\label{eso506_sim_eeuf}
\end{figure}

\begin{figure}
\centering
\includegraphics[scale=0.3]{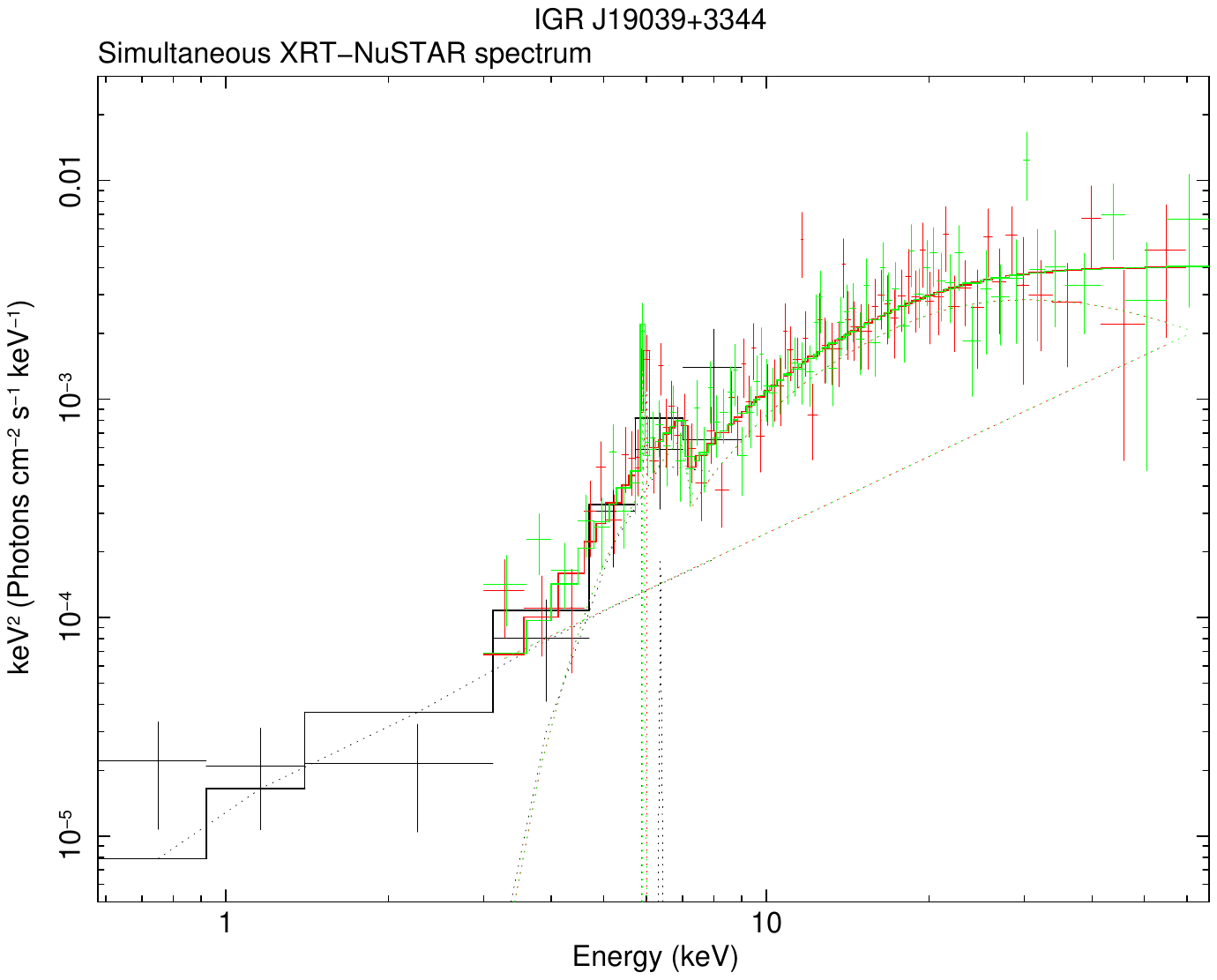}
\caption{{ Simultaneous XRT/NuSTAR unfolded spectrum of IGR J19039+3344.}}
\label{19039_sim_eeuf}
\end{figure}

\begin{figure}
\centering
\includegraphics[scale=0.3]{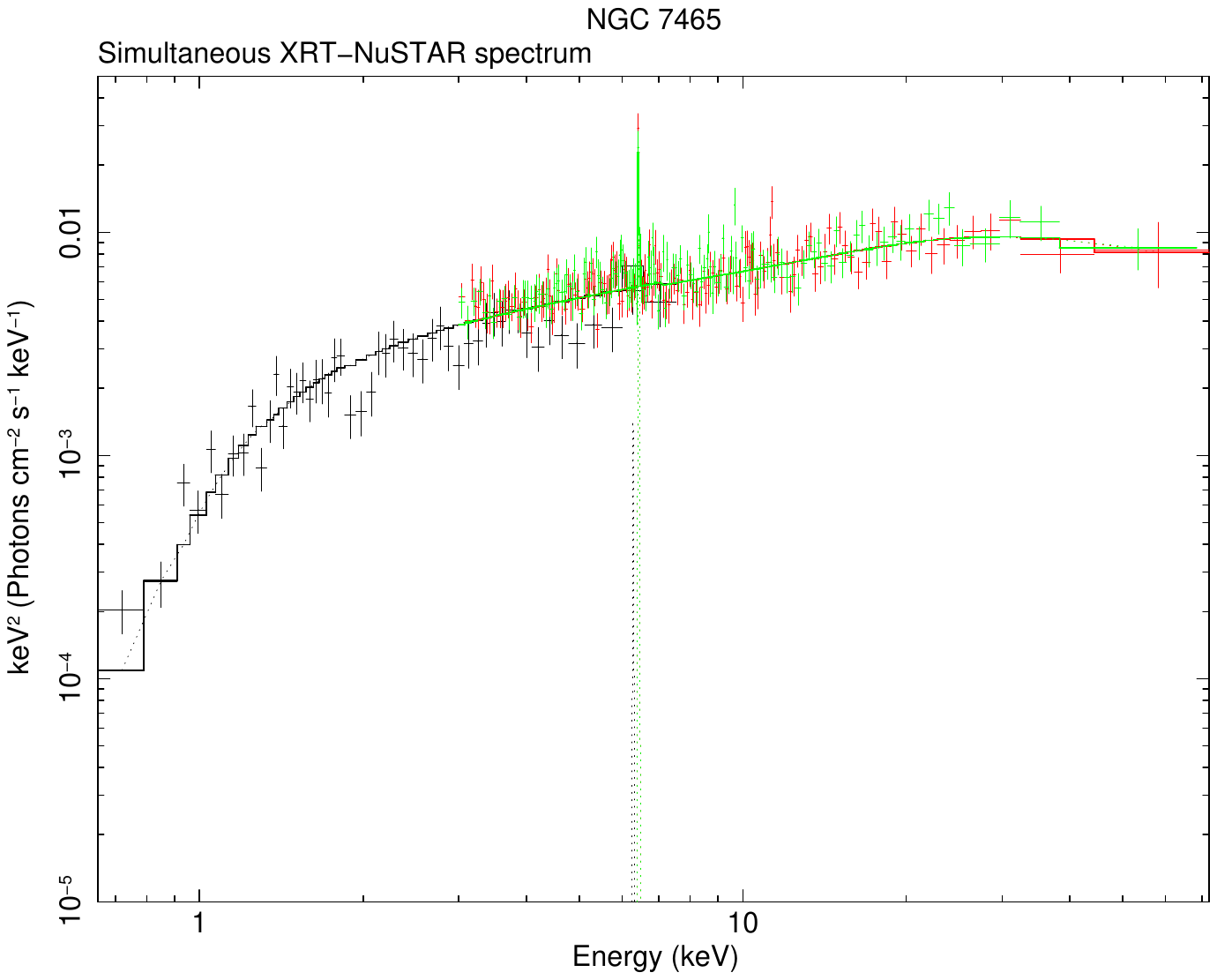}
\caption{{ Simultaneous XRT/NuSTAR unfolded spectrum of NGC 7465.}}
\label{ngc7465_sim_eeuf}
\end{figure}

As can be seen from Tables \ref{eso506_fits}, \ref{19039_fits} and \ref{ngc7465_fits},
errors on the high energy cut-off could be estimated for NGC 7465 and IGR J19039+3344, although
in the latter case they are quite large and asymmetrical, implying that the parameter is not
properly constrained. The reflection 
fraction is again constrained only in the case of NGC 7465, 
with the other two sources having only lower limits on the parameter. 
In Figures \ref{eso506_sim_eeuf}, \ref{19039_sim_eeuf} and \ref{ngc7465_sim_eeuf} we show
the unfolded spectra of the simultaneous XRT/NuSTAR data.

\subsection{Testing the \texttt{borus02} model}\label{borus}

Here we tested our simultaneous XRT/NuSTAR data against a more recent and physically motivated model,
i.e. the \texttt{borus02} model \citep{balokovic18}, which is not
particularly more complex than \texttt{pexrav} in its simplest formulation but
which accounts for a torus geometry, rather than the slab configuration assumed
by the \texttt{pexrav}. This model includes self-consistent iron line fluorescent emission lines and 
assumes the X-ray source to be at the geometric centre of the AGN,
emitting a spectrum that can be described as a phenomenological cutoff power-law, surrounded by cold and neutral gas.
This model also calculates the torus intrinsic column density, its covering factor and opening angle, as
well as the line-of-sight column density. We have used the 
\texttt{borus02} model in its simplest formulation, due to the poor quality of our data; in \texttt{XSPEC}
terminology, the model we employ is 
\texttt{c$_1$*phabs[atable(borus02\_v170323a.fits)+ zphabs*cabs*cutoffpl+c$_2$*cutoffpl+po]},

where the simple power-law approximates the soft excess when needed. We also accounted for a scattered component represented
in the model by the cutoff power-law multiplied by the constant c$_2$ (i.e. the fraction of the scattered continuum), while c$_1$ is the
cross-calibration constant between instruments.

\begin{table*}
\begin{center}
\caption{\texttt{borus02} fit results}
\renewcommand{\arraystretch}{1.2}
\resizebox{\linewidth}{!}{%
\begin{tabular}{lccccccccc}
\hline
{\bf Source}   & {\bf N$_{\rm \bf H}^{\bf\rm l.o.s.}$}&{\bf $\Gamma$}& {\bf E$_{\rm \bf cut}$}& {\bf LogN$_{\rm \bf H}^{\bf\rm torus}$}&{\bf cf$^{\bf\rm tor}$}&{\bf cos$\theta$}&{\bf $\Gamma_{\bf\rm soft}$}&{\bf c}&{\bf $\chi^{\bf 2}$ (d.o.f.)}\\ 
         &{\bf 10$^{\bf 22}$ cm$^{\bf -2}$}& &{\bf keV}&  & &&&&\\
\hline
ESO 506-G27 & 92.23$^{+14.87}_{-26.20}$ &2.02$^{+0.23}_{-0.35}$& NC& $>$22.75& 0.60$^{+0.33}_{-0.47}$&$>$0.33& 0.62 fixed&1 fixed &352.16 (357)\\
IGR J19039+3344&163.26$^{+57.89}_{-56.92}$ &1.93$^{+0.29}_{-0.26}$ &NC&NC &0.70$^{+0.04}_{-0.29}$&$>$0.42 & 0.83 fixed &1 fixed &117.19 (130)\\
NGC 7465 & 0.56$^{+0.78}_{-0.18}$& 1.74$^{+0.23}_{-0.09}$&$>$28& $>$23.77& 0.43$^{+0.37}_{-0.17}$& $>$0.41& -- &1.13$^{+0.14}_{-0.12}$&425.06 (460)\\
\hline
\end{tabular}
}
\label{borus02_fits}
\end{center}
\end{table*}
 As can be seen from Table \ref{borus02_fits}, the values we obtained for the
 line-of-sight column density are compatible with the values found using 
 our baseline model, although for ESO 506-G27 and for IGR J19039+3344 they tend
 to be a bit higher, but still compatible within errors. As far
 as the main continuum is concerned, the photon indices calculated through the
 \texttt{borus02} model are steeper than those obtained with the
 \texttt{pexrav} model, although the associated errors are quite large and therefore
 we can assume them to be compatible. The high energy cut-off is not constrained in any
 of the sources and we also found that the scattered component is negligible in all sources,
 therefore we cannot make any guesses if reflection is present or not using this model
 (for simplicity, we do not report the values of $\rm c_2$ in the table). 
 However, the \texttt{borus02} model is mainly devised to model the torus in highly absorbed 
 AGN, but also the torus parameters in our fits do not give useful information
 on this structure, confirming our claim that modelling  medium-to-low quality data with
 phenomenological models is the best approach.

\subsection{Time-averaged spectra}\label{average}
The last step in our analysis consists in modelling the time-averaged 
spectra over an even broader energy range, covering the whole 0.5-110 keV band. To do so, we have employed
spectra obtained by summing all the available XRT observations and fitted them together
with INTEGRAL/IBIS spectra. 
In order to validate the use of time-averaged spectra, we first checked whether the 
three sources analysed here are effected by flux and/or spectral variability by analysing each soft X-ray
spectrum available. 
Comparing the single XRT pointings listed in Table \ref{obs_log}, we found 
ESO 506-G27 and NGC 7465 do not show evidence of changes in their spectral shape, but only in their flux over
several observations spanning a time lapse of a few years. We are
therefore confident in using the summed XRT spectra, thus also boosting the statistics of our data. 
The same reasoning however cannot be applied to IGR J19039+3344 
(see discussion in section \ref{19039_discussion}), in which there is evidence of dramatic changes not only 
in its flux, but also in its
absorbing column density properties, and therefore spectral characteristics. 
For this reason, the source is excluded from this final
step of the analysis.

The time-averaged spectrum of ESO 506-G27 was obtained by summing all available
XRT observations (3 taken in 2005, and one each in 2018 and 2019, see Table \ref{obs_log}), thus reaching an 
exposure of more than 24 ksec (with a detection significance of more than 19$\sigma$), and then combining it with the INTEGRAL/IBIS spectrum.
Having a better soft X-ray spectrum, we were able to apply a more appropriate model to fit the  soft energy data, employing the model
\texttt{const*phabs*zxipcf(pexrav+zga+zga)} which will also be used to fit the XMM data (see section \ref{eso506_disc}). The
\texttt{zxipcf} model represents a partial covering absorption component due to partially ionized material,
and its main parameters are the column density (in units of 10$^{22}$ cm$^{-2}$) and the ionisation
of the absorbing medium. We point out that the addition of a second Gaussian line to model
the Fe K$\beta$ is required at 95\% confidence level according to the $\Delta\chi^2$ test, 
while no evidence of the Fe K edge seems to be present.
A cross-calibration constant to account for both flux variations and instrumental mismatches has also been
added and is found to be compatible with unity; results of the fit are reported in Table \ref{eso506_average}. 
Also in this case, we are not able to 
put constraints on the high energy cut-off, despite the wider energy range covered by the IBIS
data. We found a lower limit on the high energy cut-off of 93 keV which is consistent
with what generally found in the literature \citep{Malizia:2014,molina19}. Unsurprisingly, no constraint at all
is found on the reflection parameter, given the lack of spectral coverage between 10
and 25 keV. As far as the other spectral parameters are concerned, we found an agreement within their errors.

\begin{table}
\small
\begin{center}
\caption{ESO 506-G27 - Time-Averaged Spectral Fits (XRT/IBIS)}
\renewcommand{\arraystretch}{1.1}

\resizebox{\linewidth}{!}{%
\begin{tabular}{lr}
\hline
\multicolumn{2}{c}{\texttt{const*phabs*zxipcf*(pexrav+zga+zga)}}\\
\hline
 N$_{\rm H}$& (41.42$^{+19.42}_{-21.63}$)$\times$10$^{22}$cm$^{-2}$\\
 Log$\xi$&$<$0.99\\
 cf&0.998$^{+0.001}_{-0.007}$\\
 $\Gamma$&1.85$^{+0.18}_{-0.61}$\\
 E$_{\rm Fe}$ (k$\alpha$)& 6.45$\pm$0.05 keV\\
 EW& 694$^{+517}_{-331}$ eV\\
  E$_{\rm Fe}$ (k$\beta$)& 6.86$^{+0.10}_{-0.12}$ keV\\
 EW&329$^{+347}_{-214}$ eV\\
 E$_{\rm cut}$& $>$93 keV \\
 R& NC \\
 c & 0.68$^{+2.20}_{-0.38}$ \\
$\chi^{2}$ (d.o.f.)& 16.59 (21)\\  
\hline

\end{tabular}
}
\label{eso506_average}
\end{center}
\end{table}

In the case of NGC 7465, the XRT time-averaged spectrum was obtained by summing the
three available observations (one taken in 2019 and two in 2020, achieving a detection significance of
$\sim$37$\sigma$) and fit them together with
the INTEGRAL/IBIS spectrum. We employed the baseline model 
adding again a cross-calibration constant to account
for differences in fluxes and between instruments. Results are reported in the last row of
Table \ref{ngc7465_fits}; the iron line energy has been fixed to the canonical value of 6.4 keV
in order for the fit to reach convergence. 
Only a marginal agreement has been found between the time-averaged fit
and the simultaneous one. Again, in this fit the cross-calibration constant is consistent 
with being 1. The photon index is flatter than the one found in the simultaneous XRT/NuSTAR
fit (both with \texttt{pexrav} and \texttt{borus02})
and only marginally consistent (within errors). The time-averaged spectrum provides a better constraint
on the high energy cut-off, albeit its value is lower than the one obtained 
when using simultaneous data; on the other hand, the reflection fraction has only an upper
limit on its value. 

Overall, for sources which are quite faint and do not have good quality broad-band data,
both at soft and hard X-ray energies, the use of time-averaged spectra does not
improve the constraints on the fit parameters.

\section{Results on individual sources}
In the following, we discuss individual sources, highlighting the
main scientific results and comparing them with the few
pieces of information found in the literature.

\subsection{ESO 506-G27}\label{eso506_disc}
ESO 506-G27 was first identified as a high energy emitting source by \citet{tueller05},
who also proposed that the object might be a heavily absorbed AGN with an
N$_{\rm H}$ estimated to be around 10$^{23}$ cm$^{-2}$, characterised by flux 
variability on timescales of months. 
\citet{masetti06ATel} and \citet{Landi:2007} afterwards proved the 
source to be indeed a Seyfert 2 galaxy.  
Analysing the 2005 XRT observations, \citet{Landi:2007} confirmed
that the source underwent a 20\% flux change over timescales of months, but they could not properly determine the 
continuum spectral shape, due to the poor quality of the XRT data. 
Using flux ratio diagnostics, \citet{Landi:2007} place the source on the threshold between Compton-Thin
and Compton-Thick AGN (an AGN is classified as Compton-thick
if its column density exceeds 10$^{24}$cm$^{-2}$); in fact, while the L$_{\rm X}$/L$_{\rm [OIII]}$
indicates a Compton-Thick nature, the L$_{\rm X}$/L$_{\rm IR}$ and  L$_{\rm X}$/L$_{\rm HX}$
implies that the source is Compton-Thin, making ESO 506-G27 a borderline object. 
Further evidence against a Compton-Thick nature for ESO 506-G27 comes from its variability and from
the equivalent width of its iron line, which is below the expected value of $\sim$1 keV.
\citet{Winter:2008} analysed Swift/XRT and XMM-Newton data
taken in 2005 and 2006 respectively in conjunction with a time-averaged Swift/BAT spectrum;
they confirmed that ESO 506-G27 
is indeed heavily absorbed, with an estimated column density of $\sim$7$\times$10$^{23}$cm$^{-2}$ and
found to be characterised by the presence of a soft excess.
These authors also argued that the flat photon index and the relatively large iron line equivalent width could
be indicative of a Compton-Thick object, although the column density is hardly high
enough to classify the source as such. Besides, \citet{Winter:2008} do not
account for the possibility of the presence of an iron line complex,
therefore resulting in a misleading measurement of the line equivalent width.
Further confirmation of the presence of a strong 
soft excess component and of a high absorbing column density came from following studies conducted 
by employing Suzaku data (e.g. \citealt{Winter:2009} and \citealt{fukazawa11}). More 
recently also \citet{Ricci_2017}, using broad-band (again XMM and Swift/BAT) data, confirmed these
findings and furthermore characterised the source by modelling its 
spectrum with an ionised absorber; they also place constraints on both
the high energy cut-off (152 keV) and on the reflection fraction (R=0.11). However
also these authors do not investigate the iron line complex.

\begin{figure}
\centering
\includegraphics[scale=0.3]{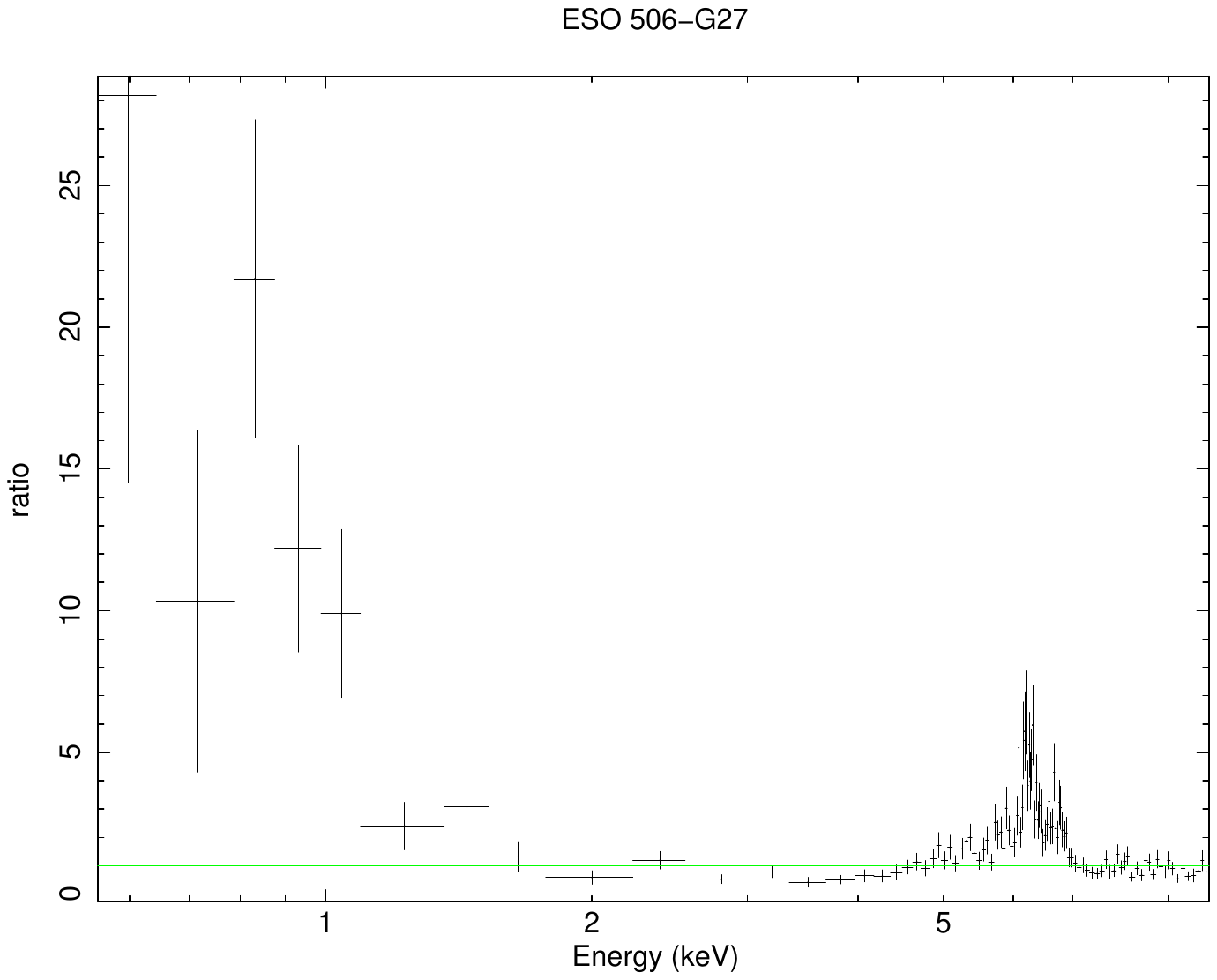}
    \caption{{XMM-Newton pn model-to-data ratio of ESO 506-G27; soft X-rays features are clearly visible below 2 keV and around 6 keV.}}
    
\label{eso506_pn_ratio}
\end{figure}

\begin{figure}
\centering
\includegraphics[scale=0.3]{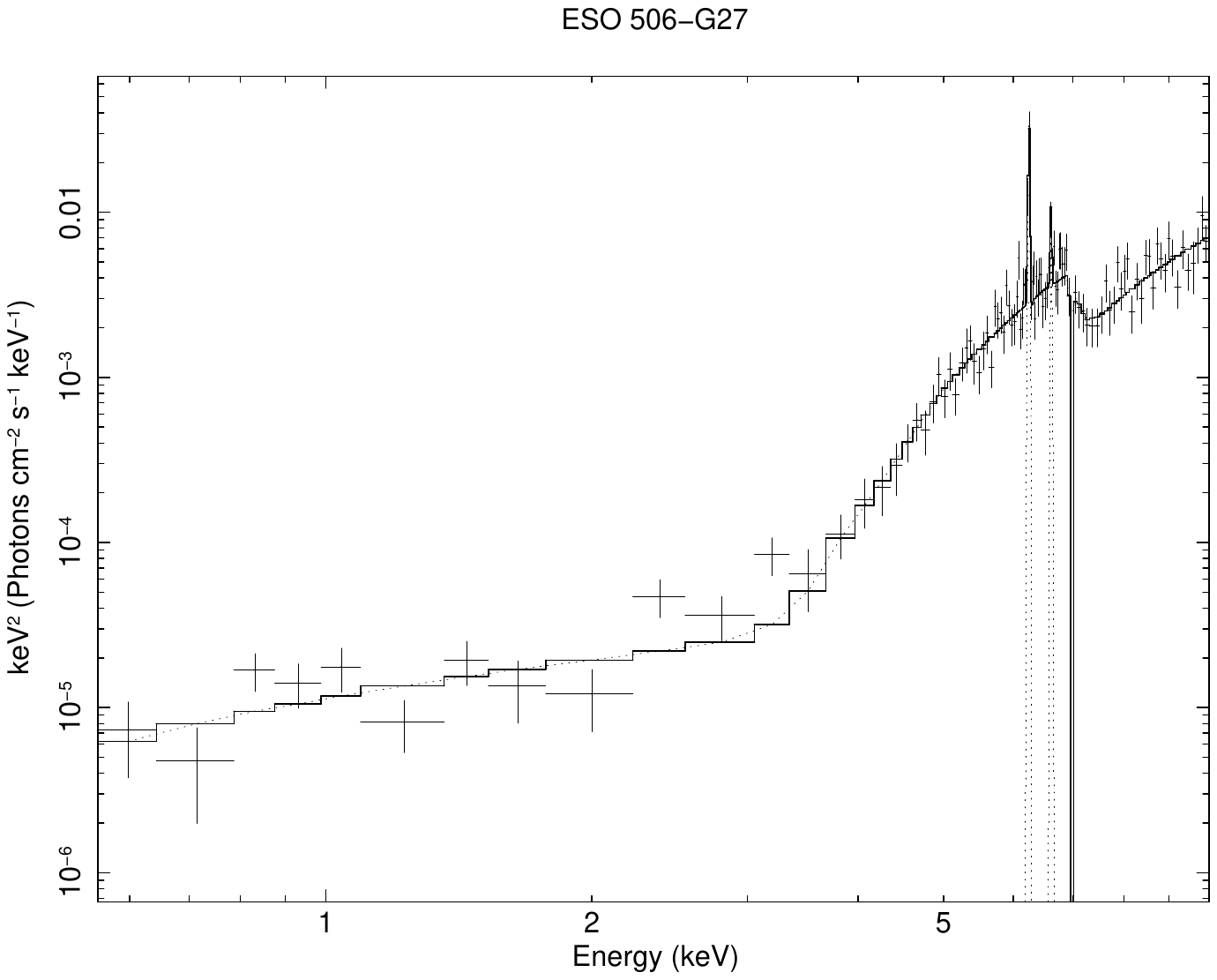}
    \caption{{XMM-Newton pn unfolded spectrum of ESO 506-G27; see text for details on spectral modelling.}}
    
\label{eso506_pn_eeuf}
\end{figure}

All these observational pieces of evidence are barely compatible with what we find in our analysis. 
We re-analysed the 2006 XMM-Newton observation of ESO 506-G27 
which has not been studied in detail before, since it is the best available data set in the soft X-ray range. 
XMM-Newton observed the source on the 24th of January, 2006
for a net total exposure of $\sim$12 ksec; here we analysed EPIC-pn data only from this 
snapshot observation. EPIC-pn \citep{Turner:2001} data were reprocessed using the 
XMM-Newton Standard Analysis Software (SAS) 
version 20.0.0 and employing the latest available calibration files. Only patterns corresponding to single
and double events (PATTERN$<$4) were taken into account and the standard selection filter FLAG=0 was applied. 
The EPIC-pn nominal exposure was filtered for periods of high background, resulting in a cleaned exposure
of $\sim$7 ksec. Source counts were extracted from a circular region of 32.5 arcsec
radius centred on the source, 
while the background spectrum was extracted from two circular regions of 20 arcsec 
radius each in source-free areas. The ancillary response matrix (ARF) and the detector response matrix 
(RMF) were generated using the XMM-SAS tasks \texttt{arfgen} and \texttt{rmfgen} and spectral channels 
were rebinned in order to achieve a minimum of 20 counts per bin.
When fitting the spectrum with a simple power-law, soft X-ray features clearly emerge, such as an
intrinsic absorption component, a soft excess below 2 keV and the Fe line complex around 6 keV
(see Fig. \ref{eso506_pn_ratio}).
We approximate the soft part of the spectrum using a partial covering absorption model in which
a partially ionized material component is also included (i.e. \texttt{zxipcf} in \texttt{XSPEC})
and introduce the k$\alpha$ and k$\beta$ iron lines. 
Our final best fit model is \texttt{phabs*zxipcf*(po+zga+zga)} (see also section
\ref{average}), which fits the data quite well, with a $\chi^2$ of 96.49 for 99 d.o.f.
(see Fig. \ref{eso506_pn_eeuf}).
We found that the source is absorbed by a mildly ionised medium
with a column density of (61.36$^{+2.51}_{-3.33}$)$\times$10$^{22}$cm$^{-2}$ and ionisation parameter
Log$\xi$=1.21$^{+0.35}_{-0.26}$, that covers almost completely the central nucleus
(cf=0.997$^{+0.001}_{-0.003}$); we confirm flux variability, since the source had a 2-10 keV
flux of $\sim$4$\times$10$^{-12}$ erg cm$^{-2}$ s$^{-1}$, twice the value measured by NuSTAR 13 years later. 
We point out that the column density we find with this model is consistent with
what we find when fitting NuSTAR data on their own, but it is only marginally consistent
with the values measured in the simultaneous XRT/NuSTAR dataset.
The photon index is 1.50$^{+0.38}_{-0.42}$, compatible within errors
with what found employing in the simultaneous XRT/NuSTAR data (see section \ref{sim_fit}). 
In the region of the iron line complex, both the
Fe K$\alpha$ and K$\beta$ lines are detected at more than 99\% confidence level according to the
$\Delta\chi^2$ test, with energies of 6.40$\pm$0.02 keV (EW=389$^{+95}_{-87}$
eV) and 6.86$^{+0.09}_{-0.10}$ keV (EW=112$^{+60}_{-62}$ eV) respectively.
All line widths are narrow and fixed to 10 eV.
Given that NuSTAR data hint at the presence of an Fe K edge, we tried to introduce
this component to the XMM data as well, by adding an absorption edge (\texttt{zedge} in
\texttt{XSPEC}) at around 7 keV. The addition of this component does not yield
any improvement in the fit, since the resulting $\chi^2$ is unchanged with respect to the
model that does not include the edge.

As a further test, we applied the same model used for NuSTAR data to the XMM spectrum,
i.e. \texttt{phabs*[po+zvfeabs*(po+zga+zga)]}. We remind that in this model, the soft excess
component is roughly approximated by a power-law, even though XMM data suggest that it 
is indeed more complex. We find a fit that is statistically equally acceptable ($\chi^2$ of 101.75 for 99
d.o.f.); in this fit we find an iron abundance slightly lower than the one found for NuSTAR data, whereas
the column density is compatible within errors. Since we cannot distinguish between these two scenarios,
a simultaneous, high-quality XMM and NuSTAR observation would be ideal to solve this ambiguity.

Our analysis confirms that ESO 506-G27 is heavily absorbed and possibly
characterised by the presence of a mildly ionised medium. 
The presence of Fe K edge detected in NuSTAR data cannot be neither rejected nor confirmed by employing
XMM data, when applying the same model. We therefore need more data to verify the significance of this feature
and investigate its true nature
However we cannot exclude a change in the intrinsic properties of the source, a likely occurrence in the
13-year span between observations; this issue will
unfortunately remain open until further and better observations of this source
are performed.
It is interesting to note that despite the use of the XMM spectrum in combination with Swift/BAT data, 
\citet{Ricci_2017} do not characterise better the iron line region complex of this source; 
nevertheless our and their broad-band fist are compatible, within the respective uncertainties, although our analysis 
is unable to put strong constraints on both reflection and cut-off energy unlike \citet{Ricci_2017}.
To overcome this limitation and confirm previous results, we also fit the XMM data together with INTEGRAL/IBIS data,
keeping in mind the source flux variability, accounted for by the introduction of a cross-calibration constant. 

When employing XMM and IBIS data together, much
in a similar way as done by \citet{Ricci_2017} with XMM and Swift/BAT data, 
we find a photon index of 1.58$^{+0.23}_{-0.32}$, but only a lower limit on the
high energy cut-off at 65 keV and no constraint at all on the reflection; the cross-calibration
constant is 0.65$^{+0.73}_{-0.23}$.

Ideally, it would have been reasonable to also fit XMM and NuSTAR data together, given
that they represent the best set of data available. However, since we cannot exclude a change
in the source physical properties between the 2006 XMM observation and the 2019 NuSTAR one,
we do not attempt such a fit.

\subsection{IGR J19039+3344} \label{19039_discussion}
IGR J19039+3344 
was first detected above 14 keV by \citet{cusumano10} (under the name 2PBC J1903.7+3349) and
then listed again in the Swift/BAT 70-month catalogue \citep{baumgartner13}. The source was  
classified as a Seyfert 2 galaxy by \citet{parisi14} and subsequently listed in the 14-year INTEGRAL/IBIS
survey catalogue by \citet{krivonos17}. 

IGR J19039+3344 is poorly studied in the X-rays, with only a non-simultaneous broad-band
spectral analysis available in the literature, based on a Swift/XRT observation taken in 2009 combined with
the average 70-month Swift/BAT spectrum \citep{Ricci_2017}. This analysis found the source to be absorbed
(log N$_{\rm H}$=22.87), with a primary power-law continuum with $\Gamma$=1.8 whose cut-off energy
could not be constrained (E$_{\rm cut}$$>$64 keV) and a possibly weak reflection component (R$<$0.5); no soft
excess component was detected. 
A part from the 2009 observation, Swift/XRT observed the source twice more in 2019 (the observation analysed in this
work) and recently in 2022; no other soft X-ray observations 
are available in the archives. 

\begin{figure}
\centering
\includegraphics[scale=0.28]{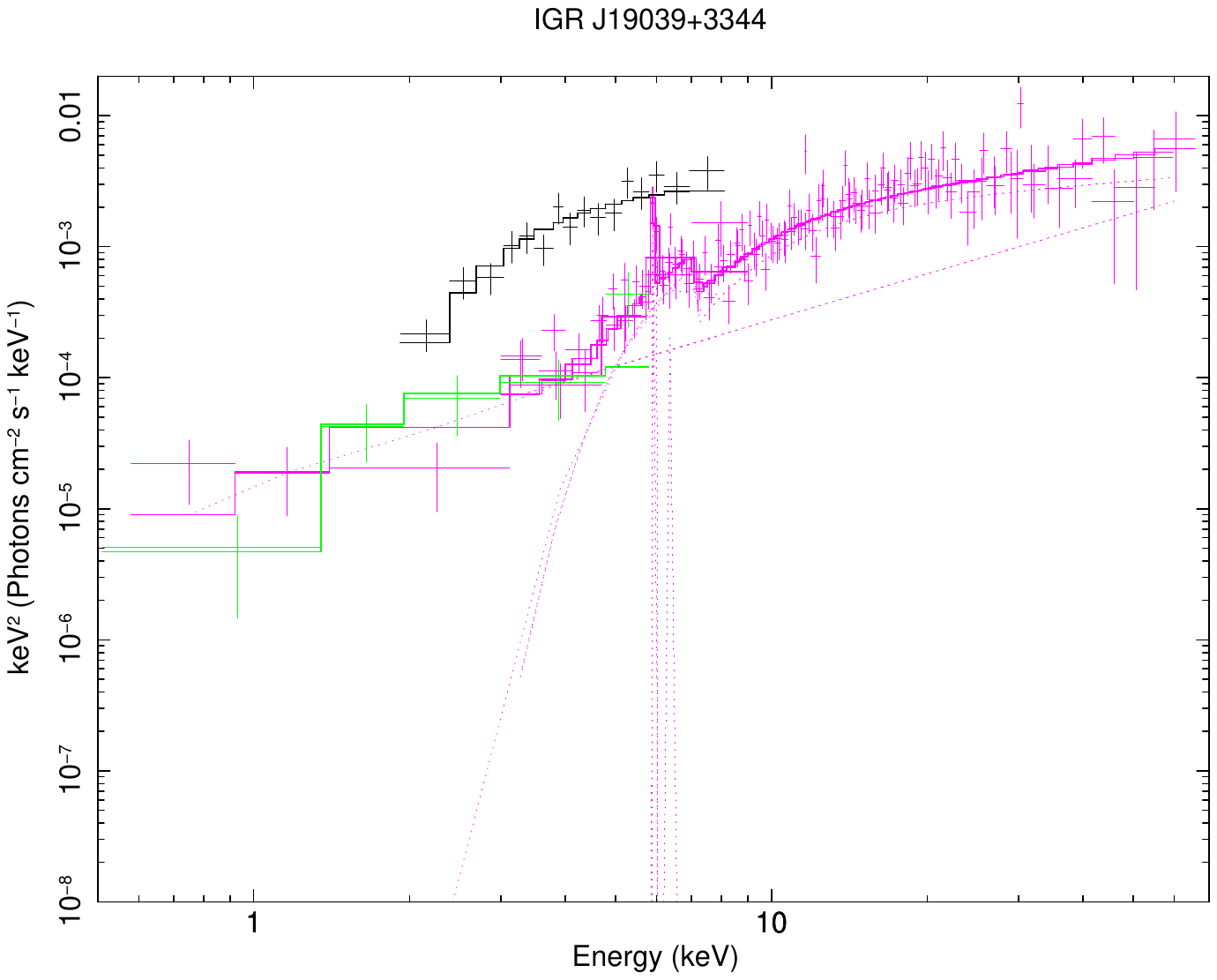}
\caption{Plot showing the three available observations fro IGR J19039+3344. In black the 2009 Swift/XRT observation is shown;
magenta represents the broad-band Swift/XRT-NuSTAR simultaneous observation obtained in 2019 and in green is
shown the latest Swift/XRT observation taken in 2022.}
\label{fig_19039}
\end{figure}

Our simultaneous broad-band spectral analysis of the 2019 observations provides marginally consistent  
results with \citet{Ricci_2017} likely due to different source states as explained later; 
given the low statistical quality of the data, neither \citet{Ricci_2017}
nor our analysis could properly determine the primary continuum shape. However, we find evidence of 
the onset of strong soft excess component,
a marginal agreement on the cut-off energy and a discrepancy on the reflection parameter.
We find a column density a factor of 5 higher 
(LogN$_{\rm H}$ = 23.7) than the one
measured in the 2009 observation, while the source 2-10 keV flux is a factor of 6 dimmer.  
These changes are even more evident when comparing single Swift/XRT snapshots (see Fig. \ref{fig_19039}): 
the 2009 observation has a column density of (6.35$^{+1.65}_{-1.19}$)$\times$10$^{22}$cm$^{-2}$ and
a 2-10 keV flux of 4.4$\times$10$^{-12}$erg cm$^{-2}$ s$^{-1}$ (in agreement with \citealt{Ricci_2017}), 
the 2019 has N$_{\rm H}$=(33.3$^{+41.0}_{-28.9}$)$\times$10$^{22}$cm$^{-2}$ and
a 2-10 keV flux of 8$\times$10$^{-13}$erg cm$^{-2}$ s$^{-1}$, while the 2022 observations has 
N$_{\rm H}$=(0.84$^{+1.75}_{-0.61}$)$\times$10$^{22}$cm$^{-2}$ with a 2-10 keV flux of 2.87$\times$10$^{-13}$erg cm$^{-2}$ s$^{-1}$,
indicating spectral and flux changes in the source at least at lower energies. All
fluxes are observed ones and not corrected for absorption.

Given this variability behaviour, combining data not strictly simultaneous is always risky: although \citet{Ricci_2017} data
relied on a XRT spectrum of higher quality than ours due to the higher flux state of the source, they are combined to a Swift/BAT
spectrum averaged over many years and possibly over different flux/spectral states; as such, their results  
must be taken with some caution. Our XRT/NuSTAR data, on the other hand, are simultaneous, but taken during a 
low flux/high column density state, which combined with the short exposures produced a lower quality spectrum
especially at low energies. As a consequence, our data do not allow a proper characterisation of the source 
broad-band spectral parameters, although we note that the column density value is now more in line with the estimated iron line EW.

Surely the most interesting result on IGR J19039+3344 is its spectral/flux change observed over timescales of years. 
The observed change in obscuration candidates the source to be a new changing look AGN \citet{ricci22}.

Changes in the line of sight absorption in the X-rays are quite a rare phenomenon to observe
(therefore difficult to probe) and are more likely to be found in samples of flux-variable AGN.
The reason behind these changes is still a matter of debate, but the most widely accepted explanation 
for changing obscuration events is related to column density variability due to clouds moving in and 
out the line-of-sight. Indeed, in the most widely accepted scenario, the torus is not considered anymore 
as uniform structure, but is rather assumed to be cloudy and/or patchy; therefore variability in the 
column density can be explained in terms of different geometry/distribution of the clouds intercepted by the line-of-sight.
However these type of changes are generally rapid, as the region involved is a small area around the nucleus. 
Given the large timescale of the variability observed in  IGR J19039+3344, alternative scenarios are also possible, 
such as changes in the ionisation state of the gas, associated with an increase or decrease of the source luminosity, 
the presence of powerful outflows or even the switch off/on of nuclear activity. In our case, we observed flux variability 
as well as absorption changes, more specifically the higher X-ray luminosity seen in 2009
could have lead to an enhancement in the ionisation state of the obscuring material, 
making it more transparent to X-ray radiation and therefore less obscured. 
The source in 2019 then returned to a lower ionisation/higher absorption state. 
However this hypothesis of changes in the ionisation properties does not explain the variation 
seen from the 2019 to the 2022 observations, when the flux was low as was the column density, although in this case  
the low statistical quality of the XRT data could have played a role.
Clearly IGR J19039+3344 deserves a long term monitoring at X-ray energies 
to confirm its changing look nature and provide more insight into the true origin of 
its changing obscuration properties.

\subsection{NGC 7465}
NGC 7465 is a LINER known to be a high energy emitter (see e.g.
\citealt{malaguti94}) and was listed in several INTEGRAL/IBIS and \emph{Swift}/BAT
catalogues. In the X-rays the source has been studied by \citet{guainazzi05}, who used Chandra
data and found the source to be characterised by a flat photon index of $\sim$0.7. These authors
model the source with two
layers of neutral material partially covering the central source, 
with N$_{\rm H}$ $\sim$ 5$\times$10$^{23}$ cm$^{-2}$ (covering 15\% of the central source) and
$\sim$2$\times$10$^{22}$ cm$^{-2}$ (covering 85\% of the central source), but opted to fix the photon
index to 2. 
These results are not fully compatible with what found in our analysis, but this could be due
to the fact that \citet{guainazzi05} made use only of narrow band data, while we took advantage of
a broad-band spectrum.
More recently, the source has been re-analysed by \citet{Ricci_2017}, employing Chandra and Swift/BAT
data;  they found  the column density to be
$\sim$3$\times$10$^{21}$ cm$^{-2}$, a quite flat photon index ($\sim$1.2) and a cut-off energy at 53 keV.
In our simultaneous broad-band analysis, we found an intrinsic column density comparable with the one reported by \citet{Ricci_2017}
but a canonical value of the photon index ($\Gamma$=1.80), exponentially declining at around 100-200 keV, plus a reflection component and associated iron line. 
The lower cut-off energy measured by both Chandra/BAT but also in our XRT/IBIS joint fit is  
likely due to the flat spectral index; given the typical AGN behaviour of the source we are more inclined to believe that 
the true spectral shape is that measured by the simultaneous broad-band XRT/NuSTAR dataset. 
According to \citet{Ricci_2018}, a high energy cutoff much above 50 keV is also expected given the source low Eddington 
ratio (Log$\lambda_{\rm Edd}$= -2.28, \citealt{koss22}).

Interestingly the X-ray absorption that we measure is 
below that estimated on the basis of ALMA data;
this is typical of type 1 unabsorbed AGN and expected if the X-ray absorption is dominated by a dust-free gas disk component at
sub-pc scales.
All these pieces of evidence suggest that NGC 7465 is a type 1 LINER, as proposed by \citet{ramos09}. 

\section{Discussion and Conclusion}
In this paper we analysed for the first time broad-band X-ray
data of three faint hard X-ray selected AGN.
These three sources (ESO 506-G27, IGR J19039+3344 and NGC 7465) have been used
as a case study to test the best way to analyse X-ray broad-band, low quality data,
in order to determine their spectral characteristics. 
The combination of the sources being faint and their low statistical quality data
make their analysis challenging and does not allow to reach conclusive results
on their intrinsic properties. This is an important issue since a large majority 
of hard X-ray selected AGN have a comparable brightness and
have only low quality X-ray data available in the archives. Most of them (mainly
sources listed in Swift-BAT catalogues and also detected by INTEGRAL/IBIS)
have indeed been simultaneously observed by Swift-XRT and NuSTAR with short pointings
and detected at a low significance level.

In the following we summarise and discuss the key findings of our study,
both from a methodological and scientific point of view.\\

{\it Methodological results.}
\begin{itemize}
    \item[-] Firstly, we have shown that in case of average luminosity AGN with poor quality X-ray spectra, 
     NuSTAR data on their own are sufficient and useful to estimate the underlying continuum emission. 
\item[-] Adding to NuSTAR spectra, data covering the soft X-ray band is preferable, since
they provide a better understanding of the physical processes at work. In particular,
simultaneous data are ideal to avoid variability issues, especially in spectral shape.
\item[-] In the case of faint sources, such as in the case of ESO 506-G27 and IGR 
J19039+3344 (see section \ref{sim_fit}), adding low quality Swift/XRT data to
average quality NuSTAR data does not however provide better results; moreover the 
cross-calibration constants had to be fixed to unity, in order to achieve
convergence in the fit.
\item[-] Time-averaged spectra, obtained by summing single snapshot
observations or resulting from long exposures taken by hard X-ray observatories
(i.e. Swift/BAT and /or INTEGRAL/IBIS), are also a useful tool if one wants to have data 
spanning over an even wider energy range. This is essential if features characterising AGN 
spectra from the soft (absorption) to the hard X-ray domain (high energy cut-off) 
are to be investigated and they might also offer a more general
picture of the sources nature. 
\item[-] However, we have shown in the present work that when the sources are quite faint,
even time-averaged spectra are not sufficient to constrain the high energy cut-off. 
\item[-] As far as spectral modelling is concerned, when the statistics are low, the use of
complex models does not necessarily imply a better description of the data. Indeed,
only in the case of NGC 7465, which is the source with the best available data, the
use of a more complex model, albeit phenomenological (i.e. \texttt{pexrav}), improves
the fit. We also point out that when dealing with low-statistics spectra, the use
of complex models can lead to an "over modelling" of the data, providing misleading
results.
\item[-] As a further test, we have demonstrated that fitting our data the more recent
and physically-motivated models such as the \texttt{borus02} does not allow a better determination
of the main spectral parameter. Indeed, important parameters such as the high energy cut-off is not constrained in any
 of the sources as well as the scattered component which is negligible and not even
 lower limits could be found. Also the torus parameters are not well constrained in all three
 sources, but the line-of-sight column density is consistent with what found using
 our baseline model.
\end{itemize}

{\it Scientific results}

\begin{itemize}
    \item[-] ESO 506-G27. Our analysis confirms that this is a heavily absorbed AGN, although not 
    Compton-Thick in nature, characterised by a strong soft excess sufficiently well-described by a mildly 
    ionised medium surrounding the central source. This result was obtained thanks archival XMM-data,
    which however are not fully consistent with more recent XRT/NuSTAR results. We report for the 
    first time, the presence of the Fe k$\beta$ line which is detected both in the XMM observation and
    in the XRT time-averaged spectrum; the feature is not detected in the NuSTAR data, likely because of the 
    inability of the detectors to separate the K$\alpha$ and K$\beta$ components.
    On the other hand, NuSTAR data show an absorption feature at $\sim$7 keV 
    possibly due to the iron K edge. This feature deserves a more in-depth 
    study  given the fact that it is not detected in the
    XMM data, suggesting spectral variability. Despite using broad-band XMM/IBIS data covering 
    the 0.5-110 keV range, we were not able to constrain the high energy cut-off and the reflection 
    fraction, in contrast with previous works (e.g. \citealt{Ricci_2017}).
    \item[-] IGR J19039+3344. This is a very peculiar source, as shown by our analysis that highlights a 
    dramatic change in its absorption properties over a 10-year span, a phenomenon which is quite rare to 
    observe. The source went from being mildly absorbed in 2009 to be heavily absorbed in 2019, as found 
    in our simultaneous Swift-XRT/NuSTAR fit, and then to unabsorbed (N$_{\rm H}$$<$10$^{22}$cm$^{-2}$)
    in 2022. Changes in observed fluxes have been reported in the analysed observations. 
    Moreover, we find evidence of a further change in its 
    spectral characteristics, with the onset of a strong soft excess component in the 2019 observation.
    Our simultaneous broad-band spectral analysis is marginally consistent with the results in 
    \citet{Ricci_2017}, particularly regarding the high energy cut-off, whereas there is a discrepancy in 
    the values of the reflection fraction. Neither \citet{Ricci_2017} nor our analysis could properly 
    determine the primary continuum shape, since we are dealing with data of very low statistical
    quality.
    \item[-] NGC 7465. Our analysis confirms that this source is a type 1 LINER, as proposed in previous 
    studies, given its low absorbing column density of the order of 
    $\sim$0.6$\times$10$^{22}$ cm$^{-2}$, contrary to
    a previous study by \citet{guainazzi05} where the N$_{\rm H}$ was found to be greater than
    10$^{23}$ cm$^{-2}$. Our 
    simultaneous broad-band spectral analysis found the source to have a photon index around 1.8, with a 
    cut-off energy located above 100 keV and a mild reflection component around 1.
    These results are only partly in agreement with \citet{Ricci_2017}; in particular, these authors
    found a much flatter photon index that might be the cause of the low value of the high energy cut-off 
    (around 50 keV). However, given that the source has a low Eddington ratio and therefore the high energy cut-
    off is expected to be at energies way above 50 keV (as is the case for our simultaneous fit), 
    we are inclined to believe that the true shape of the emission continuum of NGC 7465 is the one given
    by the XRT/NuSTAR simultaneous fit.
    
\end{itemize}

We have shown in our analysis, that having simultaneous data is not sufficient to
have a clear and complete picture of AGN, if these are faint and if the available
data are of poor quality. Indeed, we have demonstrated that when dealing with
spectra with low statistics, not only complex, more physically motivated models
cannot be safely used, but also that phenomenological models cannot
properly constrain the spectral parameters. We also point out that,
depending on which dataset is used (either simultaneous or time-averaged),
results can be very different, not only with what is found in the literature,
but also with each other; in this way, one cannot be sure of the 
robustness of the spectral results obtained. 
In conclusion
to properly characterise the spectra of faint AGN, deeper
and better data are absolutely essential, and this can only be achieved through 
dedicated observational campaigns.

\section*{Acknowledgements}
The authors acknowledge financial support from ASI
under contract n. 2019-35-HH.0. This research has made use of data and/or software provided by the High Energy Astrophysics Science Archive Research Center (HEASARC), which is a service of the Astrophysics Science Division at NASA/GSFC.
The authors wish to thank the anonymous referee for useful comments that helped improve the paper.

\section*{Data Availability}
 
The data underlying this work are available in the article. Data not specifically appearing in the 
article like some X-ray images and hard X-ray spectra will be shared on reasonable 
request by the corresponding author.



\bibliographystyle{mnras}
\bibliography{biblio} 





\bsp	
\label{lastpage}
\end{document}